\documentclass[superscriptaddress,twocolumn,showpacs,prl,floatfix]{revtex4}

\usepackage{color}
\usepackage{tabularx}
\usepackage{epsfig}
\usepackage{amsmath}

\usepackage{graphicx}

\begin{document}

\title{The Ising Spin Glass in dimension four; non-universality}

\author{P. H.~Lundow} \affiliation{Department of Mathematics and
  Mathematical Statistics, Ume{\aa} University, SE-901 87, Sweden}

\author{I. A.~Campbell} \affiliation{Laboratoire Charles Coulomb,
  Universit\'e Montpellier II, 34095 Montpellier, France}

\begin{abstract}
Extensive simulations are made on Ising Spin Glasses (ISG) with
Gaussian, Laplacian and bimodal interaction distributions in dimension
four. Standard finite size scaling analyses near and at criticality
provide estimates of the critical inverse temperatures $\beta_{c}$,
critical exponents, and critical values of a number of dimensionless
parameters. Independent estimates are obtained for $\beta_{c}$ and the
exponent $\nu$ from thermodynamic derivative peak data.  A detailed
explanation is given of scaling in the thermodynamic limit with the
ISG scaling variable $\tau = 1-\beta^2/\beta_{c}^2$ and the
appropriate scaling expressions. Data over the entire paramagnetic
range of temperatures are analysed in order to obtain further
estimates of the critical exponents together with correction to
scaling terms. The Privman-Fisher ansatz then leads to compact scaling
expressions for the whole paramagnetic regime and for all sample sizes
$L$. Comparisons between the $4$d ISG models show that the critical
dimensionless parameters characteristic of a universality class, and
the susceptibility and correlation length critical exponents $\gamma$
and $\nu$, depend on the form of the interaction distribution. From
these observations it can be deduced that critical exponents are not
universal in ISGs, at least in dimension four.

\end{abstract}


\maketitle

\section{Introduction}

The universality of critical exponents is an important and remarkably
elegant property of standard second order transitions, which has been
explored in great detail through the Renormalization Group Theory
(RGT).  The universality hypothesis states that for all systems within
a universality class the critical exponents are rigorously identical
and do not depend on the microscopic parameters of the model. However,
universality is not strictly universal; there are known \lq\lq
eccentric\rq\rq models which are exceptions and violate the
universality rule in the sense that their critical exponents vary
continuously as functions of a control variable. The most famous
example is the eight vertex model solved exactly by Baxter
\cite{baxter:71}. The exceptional physical conditions which apply in
this case were discussed in detail in Ref.~\cite{kadanoff:71}.

For Ising Spin Glasses (ISGs), the form of the interaction
distribution is a microscopic control parameter. It has been assumed
tacitly or explicitly that the members of the ISG family of
transitions obey standard universality rules, following the generally
accepted statement that \lq\lq Empirically, one finds that all systems
in nature belong to one of a comparatively small number of
universality classes\rq\rq \cite{stanley:99}. One should underline the
word \lq\lq empirically\rq\rq; we know of no formal proof that
universality must hold in ISGs. It was found thirty years ago that the
$\epsilon$-expansion for the critical exponents \cite{gardner:84} in
ISGs is not predictive since the first few orders have a
non-convergent behavior and higher orders are not known. This can be
taken as an indication that a fundamentally different theoretical
approach is required for RGT at spin glass transitions. Indeed
"classical tools of RGT analysis are not suitable for spin glasses"
\cite{parisi:01,castellana:11,angelini:13}, although no explicit
theoretical predictions have been made so far concerning universality.

ISG transition simulations are much more demanding numerically than
are those on, say, pure ferromagnet transitions with no interaction
disorder. The traditional approach in ISGs has been to study the
temperature and size dependence of observables in the near-transition
region and to estimate the critical temperature and exponents through
finite size scaling (FSS) relations after taking means over large
numbers of samples. Finite size corrections to scaling must be allowed
for explicitly which can be delicate.  From numerical data, claims of
universality have been made repeatedly for ISGs \cite{bhatt:88,
  katzgraber:06,hasenbusch:08,jorg:08} even though the estimates of
the critical exponents have varied considerably over the years (see
Ref.~\cite{katzgraber:06} for a tabulation of historic estimates).

On this FSS approach the critical inverse temperature $\beta_c$
estimates are very important for deducing reliable values for the
critical exponents \cite{notebeta}. Here we also obtain independent
estimates for $\beta_c$ and for the exponent $\nu$ through the
thermodynamic derivative peak (pseudocritical temperature) technique.
The estimates for $\beta_c$ and for $\nu$ from this analysis can be
considered to be independent of each other and the correction to
scaling only plays a minor role.  With the best estimates for
$\beta_c$ in hand the numerical data in the thermodynamic limit (ThL)
regime are then analysed with the scaling variable $\tau =
(1-\beta^2/\beta_{c}^2)$ appropriate to ISGs, together with scaling
expressions which cover the entire paramagnetic temperature regime
rather than being limited to the narrow critical region
\cite{campbell:06}.  Once values for all critical parameters have been
obtained by combining information from FSS, pseudocritical
temperature, and ThL data, through the Privman-Fisher ansatz
\cite{privman:84} compact scaling expressions can be obtained covering
the entire paramagnetic temperature range and all sizes $L$.

From a comparison of the critical values for dimensionless parameters
and for the critical exponent values, each of which is characteristic
of a universality class, we conclude that the Gaussian, Laplacian and
bimodal ISGs in dimension four are not in the same universality
class. This counter example to the general rule implies that
universality does not hold in ISGs. It is relevant that it has already
been shown experimentally that in Heisenberg spin glasses the critical
exponents depend on the strength of the Dzyaloshinski-Moriya
interaction \cite{campbell:10}.

\section{Ising Spin Glass simulations}

The Hamiltonian is as usual
\begin{equation}
  \mathcal{H}= - \sum_{ij}J_{ij}S_{i}S_{j}
  \label{ham}
\end{equation}
with the near neighbor symmetric distributions normalized to $\langle
J_{ij}^2\rangle=1$. The Ising spins live on simple hyper-cubic
lattices with periodic boundary conditions.  We have studied the
bimodal (J$4$d) model with a $\pm J$ interaction distribution,
Gaussian (G$4$d) model with a $\exp(-J^2)$ interaction distribution,
and Laplacian (L$4$d) model with a $\exp(-|J|$) interaction
distribution, all in dimension $4$.  We will compare with published
measurements on $4$d ISGs.

The simulations were carried out using the exchange Monte-Carlo method
for equilibration using so called multi-spin coding, on $2^{14}$ (up to
$L=7$) or $2^{13}$ for larger $L$ individual samples at each size. An
exchange was attempted after every sweep with a success rate of at
least 30\%. At least 40 temperatures were used forming a geometric
progression reaching down to $\beta_{\max}=0.55$ for J4d,
$\beta_{\max}=0.58$ for G4d and $\beta_{\max}= 0.65$ for L4d. This
ensures that our data span the critical temperature region which is
essential for the FSS fits. Near the critical temperature the $\beta$
step length was at most $0.004$. The various systems were deemed to
have reached equilibrium when the sample average susceptibility for
the lowest temperature showed no trend between runs. For example, for
$L=12$ this means about $200000$ sweep-exchange steps.

After equilibration, at least $200000$ measurements were made for each
sample for all sizes, taking place after every sweep-exchange step.
Data were registered for the energy $E(\beta,L)$, the correlation
length $\xi(\beta,L)$, for the spin overlap moments $\langle |q|
\rangle$, $\langle q^2\rangle$, $\langle |q|^3\rangle$, $\langle
q^4\rangle$ and the corresponding link overlap $q_{\ell}$ moments.  In
addition the correlations $\langle E(\beta,L),U(\beta,L)\rangle$
between the energy and observables $U(\beta,L)$ were also registered
so that thermodynamic derivatives could be evaluated using the
relation $\partial U(\beta,L)/\partial \beta = \langle U(\beta,L),
E(\beta,L)\rangle-\langle U(\beta,L) \rangle\langle E(\beta,L)\rangle$
where $E(\beta,L)$ is the energy \cite{ferrenberg:91}.  Bootstrap
analyses of the errors in the derivatives as well as in the
observables $U(\beta,L)$ themselves were carried out.

\section{Finite size scaling}

For the present analysis we have first observed the FSS behavior of
various dimensionless parameters, not only the familiar Binder
cumulant
\begin{equation}
  g(\beta,L)=\frac{1}{2}\left(3-\frac{[\langle q^{4}\rangle]}{[\langle q^{2}\rangle])^2}\right)
  \label{g}
\end{equation}
and the correlation length ratio $\xi(\beta,L)/L$, but also other
observables showing analogous critical behavior.  One alternative
dimensionless parameter
\begin{equation}
  W = \frac{1}{\pi-2} \left(\pi\,\frac{\langle |m|\rangle^2}{\langle m^2\rangle} - 2\right)
  \label{Wdef}
\end{equation}
was introduced in the Ising ferromagnet context in \cite{lundow:10}.
In the ISGs $m$ can be replaced by $q$ so
\begin{equation}
  W_{q} = \frac{1}{\pi-2} \left(\pi\,\frac{[\langle
    |q|\rangle]^2}{[\langle q^2\rangle]} - 2\right)
  \label{Wqdef}
\end{equation}
In the same spirit we have also introduced the dimensionless parameter
\begin{equation}
  h(\beta,L) = \frac{1}{\sqrt{\pi}-\sqrt{8}}
  \left(\sqrt{\pi}\,\frac{[\langle |q^{3}|\rangle]}{[\langle
      q^{2}\rangle]^{3/2}}-\sqrt{8}\right)
  \label{hdef}
\end{equation}
We have also registered the non-self averaging parameter $U_{22}$, the
kurtosis of the spin overlap distribution, and the moments of the
absolute spin overlap distribution, together with the variance and
kurtosis of the link overlap distribution \cite{lundow:13,lundow:12a}.
Only a fraction of these data are reported here.

Analyses with the traditional technique of estimating crossing point
temperatures $[\beta_{\mathrm{cross}},L]$, defined through
$U(\beta_{\mathrm{cross}},L)=U(\beta_{\mathrm{cross}},2L)$, have
disadvantages. The statistical errors in both sizes $L$ and $2L$
contribute to the uncertainty of the crossing temperature; the scaling
correction to the smaller size $L$ dominates and is combined with the
numerical difficulty in equilibrating at the larger size $2L$. Instead
we interpolate using the data points for each $U(\beta,L)$ so as to
obtain sets of data $U(\beta_{f},L)$ for a few fixed $\beta_{f}$ in
the critical region, after making a first rough estimate of
$\beta_{c}$. (It is important to span the range of temperatures on
both sides of the true $\beta_{c}$).  We then make a global fit with
the standard FSS expression, valid in the critical region if there is
a single dominant scaling correction term :
\begin{equation}
  U(\beta,L) = U(\beta_c,\infty) + AL^{-\omega} + B(\beta-\beta_{c})L^{1/\nu}
  \label{FSS}
\end{equation}
The fit uses all the FSS region data, and gives output estimates, with
error bars, for $\beta_c$ and the critical exponents $\nu$ and
$\omega$. The parameters of Eq.~\eqref{FSS}, and their error bars at
the 95\% level, were found by using Mathematica's built-in routines
for nonlinear model fitting. The data points for each $U$ were
obtained at some 20 fixed $\beta_{f}$ near $\beta_c$
$(0.98<\beta_{f}/\beta{c}< 1.02)$, using cubic spline
interpolation. 
Thus the six parameters are based on some 200-300 data points from
different $\beta$ and $L$. The quality of the fit is checked by
looking at not only the adjusted R-square index, which is always
extremely close to 1 in our fits. A better test is to see whether the
distribution of deviations between data and fitted model, so called
standardized residuals, can be considered zero by using Mathematica's
built-in zero-location tests, using the sign-test or T-test. The
fitted models reported here pass these tests at the 5\% level.

\section{Thermodynamic derivative peaks analysis}

The thermodynamic derivative (pseudocritical temperature) analysis can
be an efficient method for analyzing data in a ferromagnet or an ISG.
Near criticality in a ferromagnet for many observables $U$ the heights
of the peaks $D_{\max}(U,L)$ of the thermodynamic derivatives $D(U,L)
=\partial U(\beta,L)/\partial \beta$ scale for large $L$ as
\cite{ferrenberg:91,weigel:09}
\begin{equation}
  \max\left[\frac{\partial U(\beta,L)}{\partial\beta}\right]
  \propto L^{1/\nu} \left(1+ aL^{-\omega/\nu}\right)
  \label{dUdbmax}
\end{equation}
A peak height $D_{\max}(U,L)$ against $L$ log-log plot tends linearly
to $1/\nu$ at large $L$ and so provides an estimate for $\nu$ directly
without the need for a value of $\beta_{c}$ as input. Corrections to
scaling only play a minor role.

The temperature location of the derivative peak $\beta_{\max}(U,L)$
also scales as
\begin{equation}
  \beta_{c}-\beta_{\max}(L) \propto L^{-1/\nu}(1+bL^{-\omega/\nu})
  \label{betamax}
\end{equation}
so as both $\beta_{c}-\beta_{\max}(L)$ and $\max\lbrack\partial
U(\beta,L)/\partial \beta\rbrack$ vary as $L^{-1/\nu}$ at large
$L$, a plot of the peak locations against the inverse peak heights
tends linearly to $\beta_{c}$ at large $L$.  This estimate is
independent of the FSS estimate.

The observables used for $U(\beta,L)$ \cite{ferrenberg:91} can be for
instance the Binder cumulant $g(\beta,L)$, the logarithm of the finite
size susceptibility $\ln\chi(\beta,L)$, or the logarithm of the
absolute value of the magnetisation $\ln(|m|(\beta,L))$.  Each of
these data sets can give independent estimates of $\nu$ and $\beta_c$
without any initial knowledge of either parameter.  For a ferromagnet,
Ferrenberg and Landau \cite{ferrenberg:91} find this form of analysis
significantly more accurate than the standard Binder cumulant crossing
approach.


All plots of this type with different observables $U$ should
extrapolate consistently to the true $\beta_c$, with the confluent
correction only appearing as a small $L$ modification to the straight
line. Provided that the peaks for the chosen observable fall
reasonably close to $\beta_{c}$, these data are in principle much
simpler to analyse than those from the Binder crossing technique where
one must estimate simultaneously $\beta_c$ and $\nu$ together with the
strength $a$ and the exponent $\omega$ of the leading correction term.
For ISGs very much the same thermodynamic differential peak
methodology can be used as in the ferromagnet. As far as we are aware
this analysis has not been used previously in the ISG context.

\section{Thermodynamic limit scaling variables and expressions}


We will give a detailed discussion of scaling in ferromagnets and in
ISGs in the thermodynamic limit (ThL) regime, using the scaling
variable $\tau=1-\beta/\beta_{c}$ in ferromagnets and
$\tau=1-\beta^2/\beta_{c}^2$ in ISGs \cite{campbell:06}, as this
approach has been widely misunderstood or simply ignored.

The ThL regime concerns all the data for each $L$ which obey the
condition $L \gg \xi(\beta, \infty)$ where $\xi(\beta,\infty)$ is the
infinite sample size correlation length for inverse temperature
$\beta$. In this regime all observables are independent of $L$ and so
are equal to their infinite size values as measured for instance by
long series HTSE sums.  A rule of thumb for this regime is
$L/\xi(\beta,\infty) \gtrsim 6$. In order to estimate critical
exponents, an extrapolation to criticality at $\beta_c$ must complete
the overall fit to ThL data taken at finite $L$.

In the literature, numerical data on critical transition phenomena are
almost always analysed using $t=1-T/T_{c}$ as the scaling variable,
following the standard critical regime prescription of the
Renormalization Group Theory (RGT). In particular $t$ scaling has been
used in recent ISG studies
\cite{hasenbusch:08,banos:12,baity:13}. However with this variable,
explicit analyses of numerical data are limited to the finite size
scaling $L/\xi(\beta,\infty) \ll 1$ regime within a narrow critical
region in temperature; $t$ diverges at infinite temperature, so when
$t$ is used outside the critical temperature region, corrections to
scaling inevitably proliferate.

In ferromagnets, for well over fifty years the scaling variable $\tau=
1-\beta/\beta_{c}$ has been used for the analysis of the high
temperature scaling expansion (HTSE) for the susceptibility; in
particular the scaling variable $\tau$ was used in the original
discussion of confluent corrections to scaling in ferromagnets by
Wegner \cite{wegner:72} which led to the important ThL susceptibility
expression
\begin{equation}
  \chi(\beta,\infty) = C_{\chi}\tau^{-\gamma}\left(1+
  a_{\chi}\tau^{\theta} + b_{\chi}\tau + \cdots\right)
  \label{chiWeg}
\end{equation}
The first correction term is the confluent correction and the second
an analytic correction.

In the critical region $\tau$ and $t$ become equivalent, but as $\tau
\to 1$ at infinite temperature (where $\chi(\beta) \to 1$) instead of
diverging, the $\tau$ expression is well controlled over the whole
paramagnetic regime.  In principle there can be many correction terms
in Eq.~\eqref{chiWeg} but in practice to high precision a leading term
and one single further effective correction term are generally
sufficient for an analysis at the level of precision of numerical
data. This is because the leading terms in the HTSE provide strict
closure conditions on the Eq.~\eqref{chiWeg} series. Thus for the
canonical Ising ferromagnet in dimension two, there are five or more
further well identified correction terms \cite{gartenhaus:88} but in
practice with one single weak effective correction term beyond the
leading term, the expression Eq.~\eqref{chiWeg} represents the
temperature dependence of the ThL susceptibility to high precision
from criticality to infinite temperature \cite{campbell:08}. The same
remark holds for Ising ferromagnets in dimension three
\cite{campbell:11,lundow:11}. In ferromagnets the use of $t$ as the
scaling variable leads to a \lq\lq crossover\rq\rq to a high
temperature mean field behavior \cite{luijten:97}, which is a pure
artefact \cite{lundow:11}. Note that for ferromagnets with zero
temperature ordering $\tau = 1-\tanh\beta$ can be a suitable
variable \cite{katzgraber:08}.

Following a protocol well-established in ferromagnets
\cite{kouvel:64,butera:02} one can define a temperature dependent
effective exponent for the susceptibility
\begin{equation}
  \gamma(\tau)= -\frac{\partial\ln\chi(\tau)}{\partial\ln\tau}
  \label{gamtau}
\end{equation}
$\gamma(\tau)$ tends to the critical $\gamma$ as $\beta \to
\beta_{c}$, and to exactly $z\beta_{c}$ as $\beta \to 0$, where $z$ is
the number of nearest neighbors so $2d$ in simple [hyper]-cubic
lattices.

For the ThL regime $\gamma(\tau)$ can be written
\begin{equation}
  \gamma(\tau) = \gamma - \frac{a_{\chi}\theta\tau^{\theta}+b_{\chi}y\tau^{y}}
        {1+a_{\chi}\tau^{\theta}+b_{\chi}\tau^{y}}
        \label{gamtauWeg}
\end{equation}
including the leading order confluent scaling term and a further
effective higher order correction term.  The exact infinite
temperature $\tau =1$ HTSE condition on the fit parameters is
\begin{equation}
  \gamma - \frac{a_{\chi}\theta+b_{\chi}y}{1+a_{\chi}+b_{\chi}} = 2d\beta_{c}
\end{equation}
The estimates for the critical $\gamma$ and $\theta$ should be
consistent with those from FSS and from pseudocritical peak
temperature analyses.

Again in Ising ferromagnets, the analogous ThL expression for the
second moment correlation length $\xi(\tau)$ is \cite{campbell:06}
\begin{equation}
  \xi(\tau) = C_{\xi}\beta^{1/2}\tau^{-\nu}\left(1+ a_{\xi}\tau^\theta
  + \cdots\right)
  \label{xiferro}
\end{equation}
The factor $\beta^{1/2}$ arises because the generic infinite
temperature limit behavior is $\xi(\tau)/\beta^{1/2} \to 1$.  The
temperature dependent effective exponent is then
\begin{equation}
  \nu(\tau)= -\frac{\partial\ln(\xi(\tau)/\beta^{1/2})}{\partial\ln\tau}
  \label{nutauferro}
\end{equation}
so with a two correction term relation:
\begin{equation}
  \nu(\tau) = \nu - 
  \frac{a_{\xi}\theta\tau^{\theta}+b_{\xi}y\tau^{y}}
       {1+a_{\xi}\tau^{\theta}+b_{xi}\tau^{y}}
       \label{nutau}
\end{equation}
$\nu(\tau)$ tends to the critical $\nu$ as $\beta \to \beta_{c}$, and
to $d\beta_{c}$ as $\beta \to 0$ for ferromagnets on simple
[hyper]-cubic lattices.

Another effective exponent for ferromagnets is $2 - \eta(\beta) =
\partial\ln(\chi(\beta,L))/\partial\ln(\xi(\beta,L)/\beta^{1/2})$
which tends to the critical $2 - \eta(\beta_c)$ at the large $L$
critical temperature limit. Plotted against $\beta^{1/2}/\xi(\beta,L)$
the plot of this exponent is defined numerically without reference to
$\beta_c$ (as against the plots of $\gamma(\tau)$ and $\nu(\tau)$).
In the analogous ISG plots $\beta$ is replaced by $\beta^2$ and
$\beta^{1/2}$ by $\beta$, see Figs.~\ref{fig:6}, \ref{fig:13} and
\ref{fig:20}.  The extrapolation giving the estimate for
$2-\eta(\beta_{c}^2$ is independent of the estimate for $\beta_{c}$.

In ISGs, because the effective interaction energy parameter is $\langle
J^2\rangle$ and not $\langle J\rangle$, the appropriate inverse \lq\lq
temperature\rq\rq parameter is $\beta^2$ not $\beta$, so the
appropriate scaling variable is $\tau = 1-(\beta/\beta_{c})^2$, (or $w
= 1-(\tanh(\beta)/\tanh(\beta_{c}))^2$ for the bimodal ISG case).
This ISG scaling variable $\tau$ (or $w$) was used for the analysis of
the ThL ISG susceptibility immediately after the introduction of the
Edwards-Anderson ISG model \cite{singh:86, klein:91, daboul:04,
  campbell:06}. With this $\tau$ the ThL susceptibility relations
Eqs.~\eqref{chiWeg},\eqref{gamtau} and \eqref{gamtauWeg} are formally
the same as in the ferromagnet.
\begin{equation}
  \chi(\tau) = C_{\chi}\tau^{-\gamma}\left(1+ a_{\chi}\tau^\theta + \cdots\right)
  \label{chiISG}
\end{equation}
and $\gamma(\tau)$ tends to the critical $\gamma$ as $\beta^{2} \to
\beta_{c}^2$. The exact high temperature limit from HTSE is
$\gamma(\tau) \to 2d\beta_{c}^2$ as $\beta^{2} \to 0$ in simple
     [hyper]-cubic lattices.

The appropriate ISG correlation length expression is
\begin{equation}
  \xi(\tau) = C_{\xi}\beta\tau^{-\nu}\left(1+ a_{\xi}\tau^\theta + \cdots\right)
  \label{xiISG}
\end{equation}
The factor $\beta$ arises from the generic form of the ISG
$\xi(\beta)$ high temperature series \cite{campbell:06}.  The
temperature dependent ISG effective exponent is
\begin{equation}
  \nu(\tau)= - \frac{\partial\ln(\xi(\beta)/\beta)}{\partial\ln\tau}
  \label{nutauISG}
\end{equation}

We know of no HTSE calculations of the second moment correlation
length in ISGs which would lead to an exact $\beta \to 0$ limit for
$\nu(\tau)$ in ISGs.  As an empirical rule deduced from the $\chi$
HTSE analysis of Ref.~\cite{daboul:04}, in simple hyper-cubic lattices
of dimension $d$ this limit is $\nu(\beta=0)= (d-K/3)\beta_c^2$ where
$K$ is the kurtosis of the interaction distribution.

FSS analyses rely mainly on the size dependance of the critical
behavior of observables $U(\beta_c,L)$ and their derivatives
$[\partial U(\beta,L)/\partial \beta]_{\beta_c}$. For the
dimensionless parameters such as the cumulant $U_{4}(\beta,L) =
\langle m(\beta,L)^4\rangle/\langle m(\beta,L)^2\rangle^2$ or
alternatively the Binder parameter $g(\beta,L)=(3-U_{4})/2$, and the
correlation length ratio $\xi(\beta,L)/L$, the form of the critical
size dependence
\begin{equation}
  U(\beta_{c},L) = U_{\beta_{c},\infty} +
  K_{U}L^{1/\nu}\left(1+a_{U}L^{-\omega}+\cdots\right)
  \label{FSS0}
\end{equation}
and the critical derivative expression
\begin{equation}
  \left[\frac{\partial U(\beta,L)}{\partial \beta}\right]_{\beta_c} =
  K_{U^{'}}L^{1/\nu}\left(1+a_{U^{'}}L^{-\omega}+\cdots\right)
  \label{FSS1}
\end{equation}
can be retained unaltered with $\tau$ scaling for very small
$1-\beta/\beta_c$ both for ferromagnets and for ISGs.

It has been pointed out on general grounds
\cite{katzgraber:06,hasenbusch:08} that the logarithmic derivative of
the susceptibility has the form
\begin{equation}
  \frac{\partial \chi(\beta,L)/\partial \beta}{\chi(\beta,L)} =
  K_{\chi}L^{1/\nu}\left(1+a_{\chi}L^{-\omega}+\cdots\right) +K_{1}
  \label{dlnchi}
\end{equation}
No evaluation was proposed for the constant term $K_{1}$ in
Ref.~\cite{katzgraber:06,hasenbusch:08}. From the leading order $\tau$
scaling finite size expression for $\chi(\beta,L)$ \cite{campbell:06}
it is easy to show that $K_{1}= -(2-\eta)/2\beta_{c}$ in a ferromagnet
\cite{campbell:11}. In an ISG the constant term in
$(\partial\chi(\beta^2,L)/\partial\beta^2)/\chi(\beta^2,L)$ is
$K_{1}=-(2-\eta)/2\beta_{c}^2$. As pointed out in
Ref.~\cite{katzgraber:06}, for many years the published estimates of
the exponent $\nu$ in ISGs were wrong by factors of the order of $2$
because a constant $K_{1}$ term was not included in the FSS
susceptibility analyses.

As stated above, the ThL regime is limited for each $L$ by a condition
for which a rule of thumb is $ L/\xi(\beta,\infty) \gtrsim 6$, and an
extrapolation to criticality must complete the overall fit to all the
ThL data in order to estimate critical exponents.  The accuracy of
this extrapolation depends on a figure of merit, the minimum value of
$\tau$ for which the ThL condition still holds for samples of size
$L$. This figure of merit in ISGs can be taken to be
\begin{equation}
  \tau_{\min} \sim (L/6\beta_{c})^{-1/\nu}
  \label{taumin}
\end{equation}

In dimension $4$ with $\beta_{c} \sim 0.5$ and $\nu \sim 1$ the
condition implies $\tau_{\min} \sim 0.25$ if the largest size used is
$L_{m}=12$. This $\tau_{\min}$ corresponds to $\beta_{\min} \sim
0.9\beta_{c}$ so to a temperature within $10\%$ of the critical
temperature. It should be underlined that in dimension $3$ with the
appropriate parameters for ISGs, $\beta_c \sim 1$, $\nu \sim 2.5$, to
reach $\tau_{\min} \sim 0.25$ would require sample sizes to $L_{m}
\sim 200$, well beyond the maximum sizes which have been studied
numerically so far in $3$d ISGs.  When fitting to obtain the
extrapolation, no {\it a priori} assumption is made as to the value of
the dominant scaling correction exponent, which can be that of the
confluent correction (as in the $3$d ferromagnet \cite{campbell:11}),
of an analytic correction (as in the $2$d ferromagnet
\cite{campbell:08}), or of a high order effective correction if the
prefactors of the low order terms happen to be very weak. The exponent
values and the prefactor $a_{\chi}$ for the leading term are obtained
from the fit. It can be noted that an exactly equivalent procedure is
followed in the traditional fits to FSS data, where extrapolations are
made from finite $L$ to infinite size. The FSS correction exponent
$\omega$ and the ThL correction exponent $\theta$ are related by
$\omega = \theta/\nu$.  This rule provides an important consistency
test for FSS and ThL analyses on each system.  The strength of the
leading correction prefactor $a$ is an important parameter for both
FSS and ThL analyses, which is rarely quoted explicitly in
publications on numerical work on ISGs.

\section{Overall scaling plot for Ising spin glasses}

One method of showing ThL and derivative peak $\chi(\beta,L)$ data
together in ISGs (following a suggestion by K. Hukushima) is to plot
$y = \partial\beta^2/\partial\ln\chi(\beta,L)$ against $x=\beta^2$,
see Figs.~\ref{fig:1}, \ref{fig:8} and \ref{fig:15}.  (One can also
plot $y = \partial\beta^2/\partial\ln(\xi(\beta,L)/\beta)$ against
$x=\beta^2$).  These plots are purely displays of raw measured data
and do not require $\beta_{c}$ or any other parameter as input.

Each individual curve consists of data for a given $L$. Following the
derivative peak discussion above, on this plot the set of minima
points for large $L$ extrapolate linearly to the critical point $x =
\beta_{c}^2$ at $y = 0$.

The envelope curve corresponding to the data which for each $L$ are in
the ThL regime and its extrapolation to the critical point can be
fitted by an expression :
\begin{equation}
  \frac{\partial\beta^2}{\partial\ln\chi(\beta)} =
  \frac{(\beta^2-\beta_c^2)(1+a_{\chi}\tau^{\theta})} {\gamma +
    (\gamma-\theta) a_{\chi} \tau^{\theta}}
  \label{dbsqdlns}
\end{equation}
A further correction term can be readily included if needed.  The
intercept $y=0$ of the fit curve occurs at the critical point where
$x=\beta_{c}$ and the initial slope starting at the intercept is
$\partial y/\partial x =-1/\gamma$. The fit parameters are
$\beta_{c}$, $\gamma$, $\theta$ and $a_{\chi}$.  The $\chi(\beta,L)$
fit must obey the condition that at $x=0$, $y \equiv 1/2d$, so as
$\beta_c$ is known from the minima point and corrections beyond the
leading one are negligible, there are just two free fit parameters. As
a bonus, all the data used for the estimates come from temperatures
above the critical temperature, where equilibration is easier to
achieve than at and below criticality.  Thus from an analysis of
susceptibility derivative data alone, it is possible to estimate all
the principal critical parameters ($\beta_{c}, \gamma, \nu, \theta$)
for each model.


\section{Analytic corrections to scaling in ISGs}

It will be noted that the ThL data analyses presented here show no
evidence for the presence of an analytic correction term proportional
to $\tau$, which if it exists should become dominant as criticality is
approached when the confluent $\theta$ is greater than $1$.

In a ferromagnet the leading analytic term is due to the field
dependence of the analytic part of the free energy. We know of no {\it
  a priori} estimate of the strength of such terms in the ISG context,
but it is plausible that they are intrinsically weak because the field
is only present at higher order. HTSE analyses, particularly the M$1$
and M$2$ techniques \cite{daboul:04}, should be sensitive to the
presence of analytic terms.  In the HTSE measurements of Klein {\it et
  al.} \cite{klein:91} an explicit test was made for an analytic
correction in the $4$d bimodal ISG. No evidence was found for such a
term.  In all the more extensive HTSE analyses of
Ref.~\cite{daboul:04} the leading ThL correction term effective
exponent $\theta$ was always significantly greater than $1$.

It can be noted that for dimensional reasons, FSS corrections
$L^{-\omega_{i}}$ have exponents related to the equivalent ThL
exponents through $\omega_{i} = \theta_{i}/\nu$. So for ISGs in
dimension $3$ with $\nu \sim 2.5$ \cite{hasenbusch:08,baity:13}, the
leading analytic correction term would have an FSS exponent
$\omega_{a}= 1/\nu \sim 0.4$. The leading correction term estimated
from extensive $3$d bimodal ISG numerical data analysis
\cite{hasenbusch:08,baity:13} has an exponent, $\omega \sim 1.1$,
implying a dominant confluent ThL correction with exponent $\theta
\sim 2.75$. There is no mention in these publications of any analytic
FSS term with an exponent $\omega_{a} \sim 0.4$.  We conclude
empirically that quite generally in ISGs the ThL analytic correction
terms proportional to $\tau$ are small or negligible, presumably due
to vanishingly weak prefactors.

Working with the $t$ scaling variable for ISGs as in
\cite{hasenbusch:08,banos:12} means that information on critical
exponents coming from data temperatures well above criticality is
lost.  Comments which have been published such as "The difference
between the [$\tau$ scaling] expressions and the standard expressions
is only in the corrections to scaling." \cite{katzgraber:06} or "[The
  $\tau$ scaling] approach might partly take into account the scaling
corrections..."  \cite{hasenbusch:08} are incorrect and follow from a
misunderstanding.  They refer to the initial lowest order form of
$\tau$ scaling \cite{campbell:06} where corrections to scaling were
explicitly left out of the analysis. Full expressions including the
Wegner correction terms have been used in the analysis of ferromagnets
\cite{campbell:08,campbell:11,lundow:11} and are used here for
ISGs. It is helpful to note that because of exact infinite temperature
closure conditions on $\chi(\tau)$, $\xi(\tau)/\beta$, $\gamma(\tau)$
and $\nu(\tau)$ from HTSE, a potentially infinite set of high
temperature corrections can generally be grouped together into a
single effective correction.

\section{Privman-Fisher scaling}

The Privman-Fisher FSS ansatz Eq.~\eqref{PFchi} was originally
presented \cite{privman:84} in terms of scaling near criticality, with
$t$ as the scaling variable. With the $\tau$ scaling expressions the
ansatz can be applied successfully over the entire paramagnetic
temperature range (see Ref.~\cite{campbell:11} for a ferromagnet
case).  Once estimated values for $\beta_{c},\gamma, \nu, \theta,
a_{\chi}, a_{\xi}$ (and possible higher order correction terms if
necessary) have been obtained by fits to the ThL regime and
extrapolations to criticality, one has explicit expressions for
$\chi(\tau, \infty)$ and $\xi(\tau,\infty)$ for the entire
paramagnetic regime. Privman-Fisher ansatz \cite{privman:84} scaling
plots then can be made for all $L$ and all the paramagnetic regime :
\begin{equation}
  \frac{\chi(\tau,L)}{\chi(\tau,\infty)} = F_{\chi}[L/\xi(\tau,\infty)] +
  \frac{a_{(\omega,\chi)}}{L^{\omega}}G_{\chi}[L/\xi(\tau,\infty)]
  \label{PFchi}
\end{equation}
and
\begin{equation}
  \frac{\xi(\tau,L)}{\xi(\tau,\infty)} = F_{\xi}[L/\xi(\tau,\infty)] +
  \frac{a_{(\omega,\xi)}}{L^{\omega}}G_{\xi}[L/\xi(\tau,\infty)]
  \label{PFxi}
\end{equation}
Scaling, with $\omega = \theta/\nu$, should be "perfect" for all
$L$ and over the whole paramagnetic temperature range including the
critical FSS regime if the critical parameter estimates have been
chosen correctly. This overall scaling can be considered to provide an
ultimate validation of the coherence of ThL and FSS fits.

In the case of the Privman-Fisher procedure with $\tau$ scaling
applied to the cubic Ising ferromagnet susceptibility
\cite{campbell:11}, a simple explicit form for the principal
Privman-Fisher scaling function was proposed as a further ansatz:
\begin{equation}
  F_{\chi}(x) = \left(1-\exp(-bx^{(2-\eta)/a})\right)^{a}
  \label{PFansatz}
\end{equation}
where $x = L/\xi(\beta,\infty)$, and $a$ and $b$ are fit
parameters. This extremely compact form automatically fulfils the
limit conditions for large and small $x$. If this ansatz turns out to
be generally applicable, the parameters $a$ and $b$ as well as $\eta$
should be characteristics of a universality class.  For the
$\xi(\beta,L)$ scaling plot the fit ansatz becomes even simpler :
\begin{equation}
  F_{\xi}(x) = \left(1-\exp(-bx^{1/a})\right)^{a}
  \label{PFxiansatz}
\end{equation}
with different $a$ and $b$ parameters.  The same approach will be
applied below to the ISGs in dimension four.

\section{The Gaussian ISG in dimension $4$}

We will now address the question of specific ISGs in dimension four,
starting with the Gaussian interaction distribution.  Simulation
measurements up to $L=10$ were published on the $4$d Gaussian ISG,
together with a $4$d bimodal ISG with diluted interactions ($65\%$ of
the interactions being set to $J=0$) \cite{jorg:08}.  The critical
temperature for the 4d Gaussian ISG was estimated from Binder
parameter and correlation length ratio measurements to be $T_{c}=
1.805(10)$ so $\beta_{c} = 0.554(3)$, in agreement with earlier
simulation estimates $0.555(3)$ \cite{parisi:96,ney:98} and consistent
with a high temperature series expansion (HTSE) estimate $\beta_{c}^2
= 0.314(4)$, i.e. $\beta_{c}= 0.560(4)$ \cite{daboul:04}. The
simulation analyses \cite{jorg:08} led to essentially identical
exponents for the two systems : $\nu =1.02(2)$ and $\eta = -0.275(25)$
and so through scaling rules $\gamma = 2.32(8)$ \cite{jorg:08}. The
HTSE critical exponent estimates were $\gamma= 2.3(1)$ and $\theta
\sim 1.35$ \cite{daboul:04}. The two systems of Ref.~\cite{jorg:08}
happen to show small (for the Gaussian) and almost negligible (for the
diluted bimodal) corrections to scaling for the Binder parameter,
rendering the $\beta_{c}$ estimates particularly reliable.

We have repeated the Gaussian measurements of the Binder parameter
$g(\beta,L)$ and the correlation length ratio $\xi(\beta,L)/L$, and
have also measured the dimensionless parameters $W_{q}(\beta,L)$,
Eq.~\eqref{Wqdef}, and $h(\beta,L)$, Eq.~\eqref{hdef}, in the critical
region.  Plots of $W_{q}(\beta,L)$ for chosen fixed $\beta$ as
functions of $L^{-\omega}$ for a fixed $\omega$ are shown in
Fig.~\ref{fig:2}, together with fits as described above,
Eq.~\eqref{FSS}. Due to the weakness of the corrections to scaling for
Gaussian interactions ($A$ in Table~\ref{Table:I}) choosing $\omega$
to be $1.5$, $2.0$ or $2.5$ made little difference to the output
optimal fit parameters. The figures for the other dimensionless
parameters are similar and are not presented explicitly for space
considerations.  The overall fit parameters including uncertainties
due to $\omega$ are given in Table~\ref{Table:I}.





\begin{table}[htbp]
  \caption{\label{Table:I} Values of the $4$d Gaussian ISG FSS
    analysis fit parameters with standard errors. Dimensionless
    parameters $g(\beta,L)$, $W_{q}(\beta,L)$, $h(\beta,L)$ and
    $[\xi/L](\beta,L)$.}
  \begin{ruledtabular}
    \begin{tabular}{ccc}
      $g(\beta_c,\infty)$&$0.484$&$0.003$ \\
      $A(g)$&$0.09$&$0.04$\\
      $B(g)$&$0.6028$&$0.016$\\
      $\beta_{c}(g)$&$0.5571$&$0.0004$\\
      $\nu(g)$&$1.024$&$0.013$\\
      $W_{q}(\beta_c,\infty)$&$0.248$&$0.003$ \\
      $A(W_{q})$&$0.13$&$0.03$\\
      $B(W_{q})$&$0.471$&$0.009$\\
      $\beta_{c}(W_{q})$&$0.5565$&$0.0005$\\
      $\nu(W_{q})$&$1.029$&$0.009$\\
      $h(\beta_c,\infty)$&$0.394$&$0.004$ \\
      $A(h)$&$0.11$&$0.04$\\
      $B(h)$&$0.578$&$0.013$\\
      $\beta_{c}(h)$&$0.5570$&$0.0004$\\
      $\nu(h)$&$1.029$&$0.015$\\
      $[\xi/L](\beta_c,\infty)$&$0.451$&$0.006$ \\
      $A(\xi/L)$&$0.3$&$0.1$\\
      $B(\xi/L)$&$0.447$&$0.009$\\
      $\beta_{c}(\xi/L)$&$0.5565$&$0.0010$\\
      $\nu(\xi/L)$&$1.03$&$0.01$\\
    \end{tabular}
  \end{ruledtabular}
\end{table}

We thus obtain consistent estimates, Table~\ref{Table:I}, $\beta_{c}=
0.557(1)$, $\nu = 1.029(5)$ (taking the average values) together with
the infinite size limit dimensionless critical parameter values
$g(\beta_c,\infty) = 0.484(3)$, $W_{q}(\beta_{c},\infty)=0.2418(3)$,
$h(\beta_{c},\infty)=0.394(4)$ and
$[\xi/L](\beta_c,\infty)=0.451(6)$. The estimates from the Binder
cumulant are in agreement with those of Ref.~\cite{jorg:08} where
$g(\beta_c,\infty) = 0.470(5)$ (no value for $[\xi/L](\beta_c,\infty)$
was cited explicitly and $W_{q}(\beta,L)$, $h(\beta,L)$ were not
measured). The present $\beta_{c}$ value is rather more accurate
mainly because of a more closely spaced temperature grid and better
statistics to higher $L$.  The FSS scaling rule at criticality is
$\chi(\beta_c,L) \propto L^{2-\eta}$.  Using the present $\beta_{c}$
estimate a FSS log-log plot of $\chi(\beta_{c},L)/L^2$ against $L$
gives a straight line of slope $-\eta = 0.307(10)$, consistent with
the estimate $-\eta= 0.275(25)$ of \cite{jorg:08}. From the scaling
rule $\gamma = (2-\eta)\nu$ we obtain a FSS estimate $\gamma =
2.35(1)$.

Thermodynamic derivative peak data are shown in the form of peak
location $\beta_{\max}(L)$ against inverse peak height $1/D_{\max}(L)$
for the derivatives $\partial g(\beta,L) \partial \beta$, $\partial
W_{q}(\beta,L)/\partial\beta$, and $\partial h(\beta,L) \partial
\beta$, Fig.~\ref{fig:3}. The linear extrapolations to
$1/D_{\max}(L)=0$ lead consistently to $\beta_{c} = 0.557$, in full
agreement with the FSS estimate.  Log-log plots of $D_{\max}(L)$
against $L$ have limiting slopes of $0.99(1)$, so from the scaling
rule Eq.~\eqref{dUdbmax} $\nu= 1/0.99(1)=1.01(1)$ again consistent
with the FSS estimate.  This $\nu$ estimate is independent of the
estimate for $\beta_{c}$.

The temperature dependent effective exponents $\gamma(\tau,L)$ and
$\nu(\tau,L)$, Eqs.~\eqref{gamtau}, \eqref{nutauISG} assuming
$\beta_{c}=0.557$ are shown in Figs.~\ref{fig:4} and \ref{fig:5}.  The
ThL regime can be recognized by the condition $\gamma(\tau,L)$ or
$\nu(\tau,L)$ becoming independent of $L$, or from the figure of merit
Eq.~\eqref{taumin}.  The ThL regime correlation length exponent
$\nu(\tau)$ has only a weak temperature variation. The overall fit to
the $\nu(\tau,L)$ data in Fig.~\ref{fig:5} gives a ThL temperature
dependent exponent
\begin{eqnarray}
  \nu(\tau) =
  1.032-\frac{0.041\cdot 1.6\, \tau^{1.6}+0.017\cdot 3\, \tau^3}
  {1+0.041\, \tau^{1.6}+0.017\, \tau^3}
\end{eqnarray}
or
\begin{equation}
  \xi(\beta,\infty)= 0.95\beta\tau^{-1.032}
  \left(1+0.041\,\tau^{1.6}+0.017\,\tau^3\right)
\end{equation}
so with $\nu=1.032(5),\theta \sim 1.6, a_{\xi}=0.041(5)$ and a weak
higher order contribution.  The correction exponent estimate $\theta
\sim 1.6$ is compatible with the HTSE value $\theta \sim 1.35$
\cite{daboul:04}.

\begin{figure}
  \includegraphics[width=3.5in]{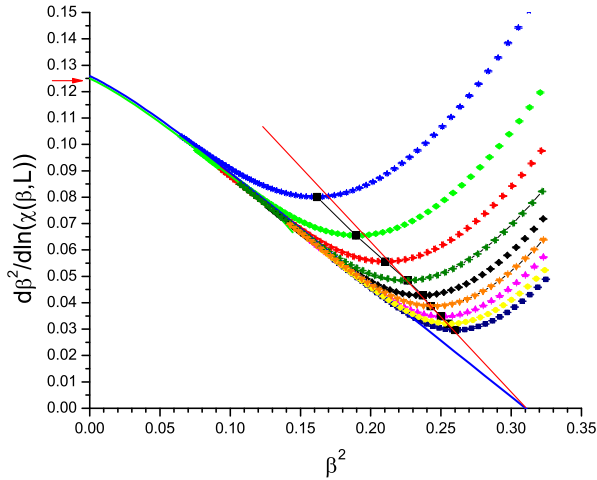}
  \caption{(Color online) The $4$d Gaussian derivatives $\partial
    \beta^2/ \partial \ln\chi(\beta,L)$ as a function of $\beta^2$.
    Throughout the paper, the symbol code for sizes is : $L=14$ purple
    pentagons, $L=13$ wine stars, $L=12$ navy squares, $L=11$ yellow
    hexagons, $L10$ pink circles,$L=9$ orange balls, $L=8$ black
    triangles, $L=7$ olive inverted triangles, $L=6$ red diamonds,
    $L=5$ green left triangles, $L=4$ blue right triangles. (For
    display purposes not all sizes are shown in each Figure).  Blue
    curve : ThL regime data fit.  Large squares : minima locations for
    each $L$, straight line : extrapolation to
    $\beta_{c}^2$. }\protect\label{fig:1}
\end{figure}

\begin{figure}
  \includegraphics[width=3.5in]{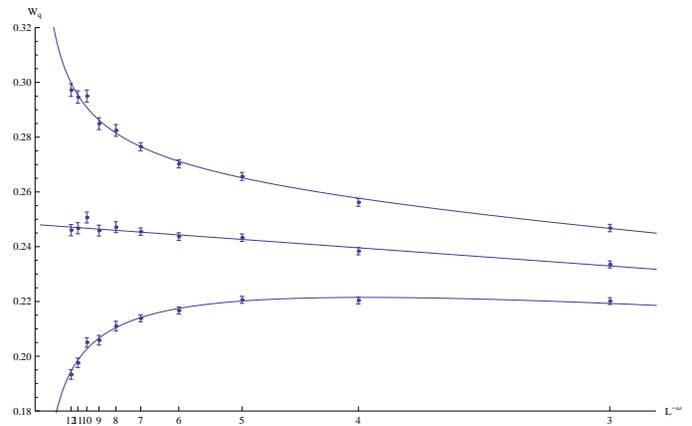}
  \caption{(Color online) The $4$d Gaussian $W_{q}$ finite size
    scaling. Data and fit curves : upper $\beta=0.5665$, centre
    $\beta=0.5565$, lower $\beta=0.5465$.  Fit value $\omega = 2$. See
    text and Table~\ref{Table:I}. }\protect\label{fig:2}
\end{figure}

\begin{figure}
  \includegraphics[width=3.5in]{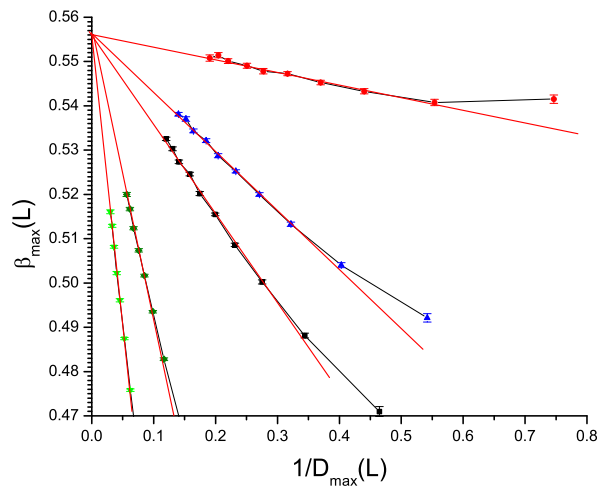}
  \caption{(Color online) The $4$d Gaussian thermodynamic derivative
    peaks; peak location $\beta_{\max}(L)$ as function of inverse peak
    heights $1/D(L)$ where $D(L) = \max \partial U(\beta,L)/ \partial
      \beta$.  Observables $U$: red circles $W_{q}$, blue
    triangles $h$, black squares $g$, green inverted triangles
    $\ln\chi$, olive diamonds $\ln |q|$. Straight line fits indicate
    extrapolations to $\beta_c$ (see text).}\protect\label{fig:3}
\end{figure}

\begin{figure}
  \includegraphics[width=3.5in]{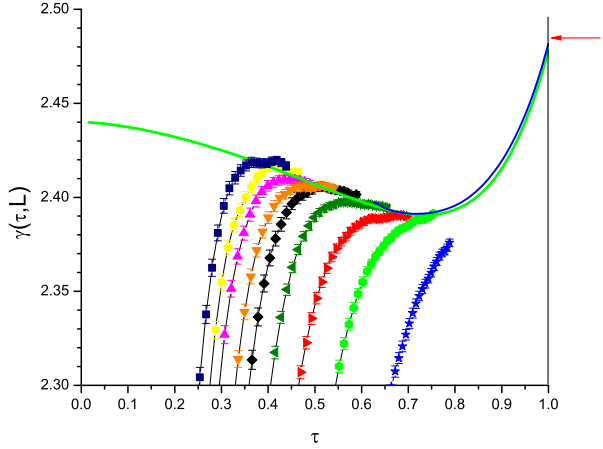}
  \caption{(Color online) The $4$d Gaussian temperature dependent
    effective exponent $\gamma(\tau)=\partial \chi(\beta,L)/ \partial
    \tau$ for $\beta_{c}= 0.557$. Symbol code as in Fig.~\ref{fig:1}. Blue
    curve : HTSE data points. Green curve : ThL regime data
    fit. }\protect\label{fig:4}
\end{figure}

\begin{figure}
  \includegraphics[width=3.5in]{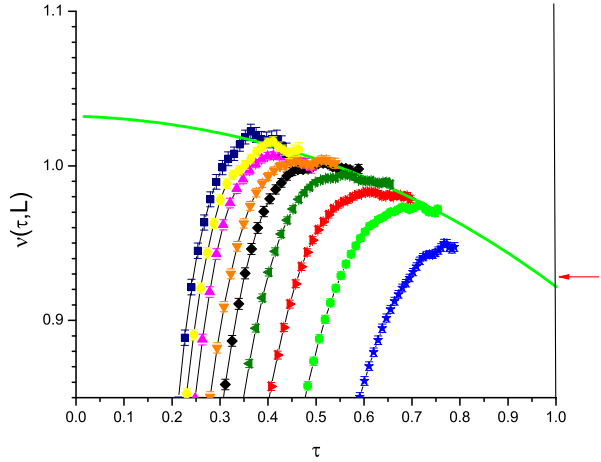}
  \caption{(Color online) The $4$d Gaussian effective
    $\nu(\tau)=\partial (\xi(\beta,L)/\beta)/ \partial \tau$ for
    $\beta_{c}= 0.557$. Symbol code as in Fig.~\ref{fig:1}. Blue curve : ThL
    regime data fit. }\protect\label{fig:5}
\end{figure}

\begin{figure}
  \includegraphics[width=3.5in]{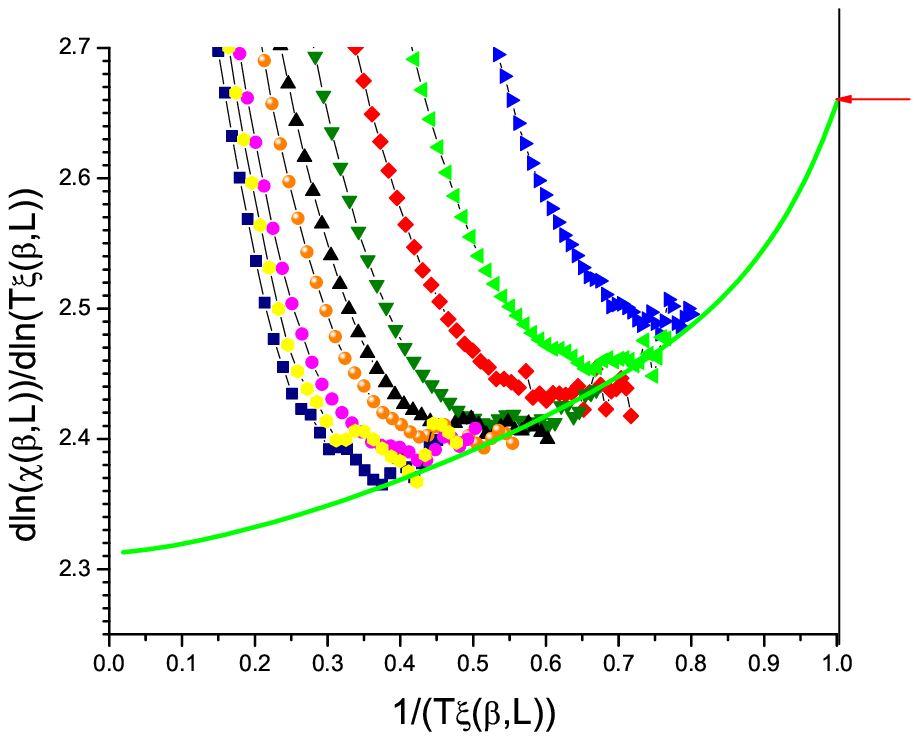}
  \caption{(Color online) The $4$d Gaussian effective exponent
    $2-\eta(\beta)=\partial\ln\chi(\beta,L)/\partial\ln(\xi(\beta,L)/\beta)$
    against $\beta/\xi(\beta,L)$.  Symbol code as in
    Fig.~\ref{fig:1}. Green envelope curve : ThL regime data fit and
    extrapolation. }\protect\label{fig:6}
\end{figure}

The $\gamma(\tau)$ curve evaluated directly from the HTSE series
\cite{daboul:04} is exact at high and moderate temperatures, once
$\beta_c$ is estimated, and is fully consistent with the ThL
$\gamma(\tau,L)$ simulation data in the appropriate $\tau$ range,
Fig.~\ref{fig:4} (Fig.~\ref{fig:1} and Fig.~\ref{fig:4} are
alternative presentations of the same derivative data).  The
simulation and HTSE data taken together show that for the leading
confluent correction term $\tau(\theta)$ the prefactor is small, and
that there is a weak high order effective correction term, so
\begin{eqnarray}
  \gamma(\tau) =
  2.44-\frac{0.06\cdot 1.6\,\tau^{1.6}-0.017\cdot 8\,\tau^8}
  {1+0.06\,\tau^{1.6}-0.017\,\tau^8}
\end{eqnarray}
i.e.
\begin{equation}
  \chi(\beta,\infty)= 0.96\,\tau^{-2.44}
  \left(1+0.06\,\tau^{1.6}-0.017\,\tau^8\right)
\end{equation}
Hence $\gamma= 2.44(2)$, $\theta_{\mathrm{eff}}\sim 1.6$ and
$a_{\chi}= 0.06(1)$.  From the extrapolated derivative
$\partial\ln\chi(\beta,L)/\partial\ln(\xi(\beta,L)/\beta)$,
Fig.~\ref{fig:6}, $\eta = -0.36(4)$. This estimate is independent of
the estimate for $\beta_{c}$.

These critical exponents : $\nu=1.032(5)$, $\gamma=2.43(2)$,
$\eta=-0.36(4)$ estimated from the ThL data are thus in full agreement
with the FSS estimates above, and with the slightly less precise FSS
estimations of Ref.~\cite{jorg:08} : $\nu =1.02(2)$, $\gamma=2.32(8)$,
$\eta = -0.275(25)$. The observed exponents $\gamma=2.43(2)$ and
$\theta \sim 1.6$ are also consistent with the HTSE estimates $\gamma=
2.3(1)$ and $\theta \sim 1.35$ \cite{daboul:04}.  There is thus
excellent overall consistency.  The critical temperature and exponent
values are particularly reliable in this model because of the
accidental weakness of the correction to scaling terms.

\begin{figure}
  \includegraphics[width=3.5in]{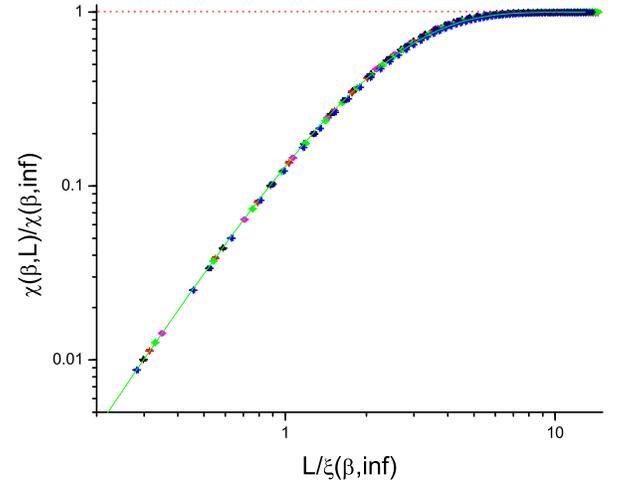}
  \caption{(Color online) The $4$d Gaussian Privman-Fisher $\chi$ FSS
    plot, $\chi(\tau,L)/\chi(\tau,\infty)$ against
    $L/\xi(\tau,\infty)$ where $\chi(\tau,\infty)$ and
    $\xi(\tau,\infty)$ are fit and extrapolation values from the
    $\gamma(\tau)$ and $\nu(\tau)$ analyses. Symbol code as in
    Fig.~\ref{fig:1}. Curve : fit (see text)}\protect\label{fig:7}
\end{figure}

\begin{figure}
  \includegraphics[width=3.5in]{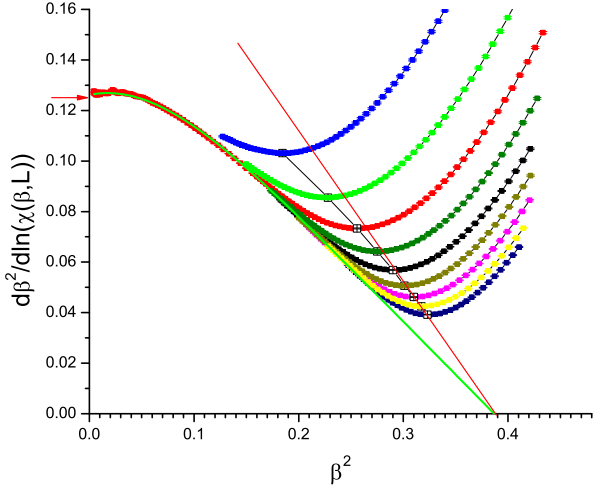}
  \caption{(Color online) The $4$d Laplacian derivatives $\partial
    \beta^2/ \partial \ln\chi(\beta,L)$ as a function of $\beta^2$.
    Throughout the paper, the symbol code for sizes is as in
    Fig.~\ref{fig:1}.  Blue curve : ThL regime data fit. Green curve:
    ThL regime envelope fit.  Large squares : minima locations for
    each $L$. Straight line : extrapolation to
    $\beta_{c}^2$. }\protect\label{fig:8}
\end{figure}

With the ThL $\chi(\beta,\infty)$ and $\xi(\beta,\infty)$ expressions
in hand we make up the Privman-Fisher susceptibility plot
$\chi(\beta,L)/\chi(\beta,\infty)= F_{\chi}[L/\xi(\beta,\infty)]$,
Fig.~\ref{fig:7}. The whole data set for the entire paramagnetic temperature
region and all $L$ (so covering the FSS regime, the ThL regime, and
the intermediate regime) shows an excellent scaling, which can be
fitted by the ansatz Eq.~\eqref{PFansatz} using the fit parameters
$\eta=-0.36(3)$, $a= 1.89(2)$, $b =0.43(2)$. The Privman-Fisher
correlation length plot can be fitted with parameters $a=0.80(2)$ and
$b=0.35(2)$ in Eq.~\eqref{PFxiansatz}.  There is a small finite size
scaling correction which we have not attempted to analyse.

We have no data for the diluted bimodal ISG of Ref.~\cite{jorg:08} but
in view of the fact that the FSS data in that model displayed the
figures show even weaker corrections to scaling than for the Gaussian,
the exponent and dimensionless parameter estimates $\nu=1.025(15)$,
$\gamma=2.33(6)$, $\eta=-0.275(25)$, $g(\beta_{c})=0.472(2)$ are
certainly very reliable also.

\section{The Laplacian interaction ISG in dimension $4$}

\begin{figure}
  \includegraphics[width=3.5in]{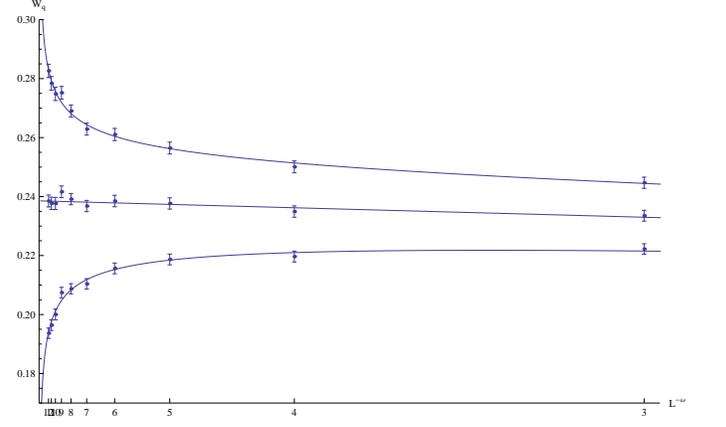}
  \caption{(Color online) The $4$d Laplacian $W_{q}$ finite size
    scaling. Data and fit curves : upper $\beta=0.6315$, centre
    $\beta=0.6215$, lower $\beta=0.6115$.  Fit value
    $\omega = 3$. See text and
    Table~\ref{Table:II}. }\protect\label{fig:9}
\end{figure}

\begin{figure}
  \includegraphics[width=3.5in]{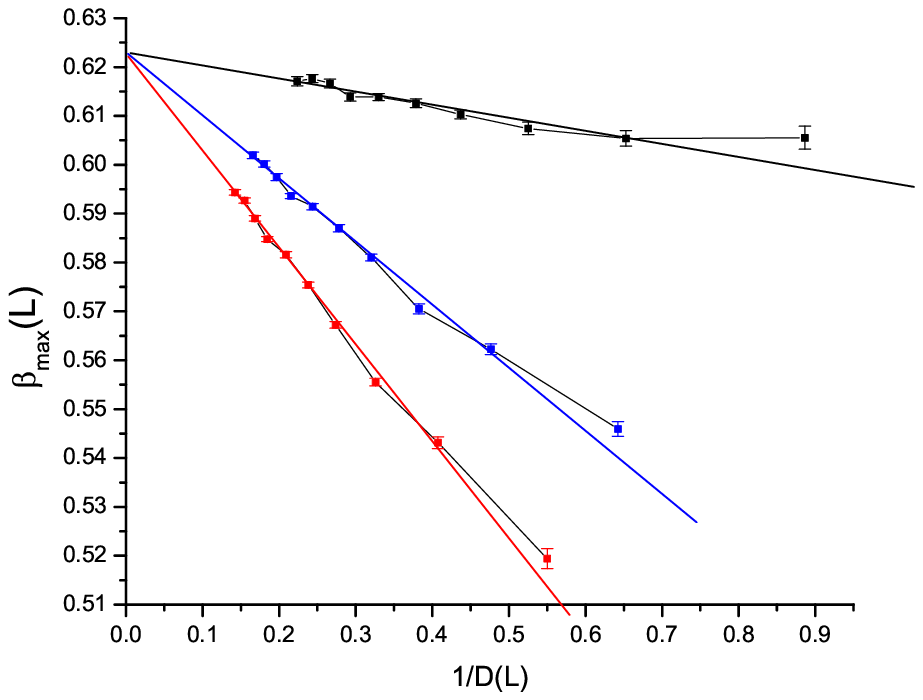}
  \caption{(Color online) The $4$d Laplacian thermodynamic derivative
    peaks; peak location $\beta_{\max}(L)$ as function of inverse peak
    heights $1/D_{\max}(L)$ where $D_{\max}(L) =\max \partial U(\beta,L)/
      \partial \beta$.  Observables $U$: red circles $W_{q}$,
    blue triangles $h$, black squares $g$, Straight line fits indicate
    extrapolations to $\beta_c$ (see text).}\protect\label{fig:10}
\end{figure}

\begin{figure}
  \includegraphics[width=3.5in]{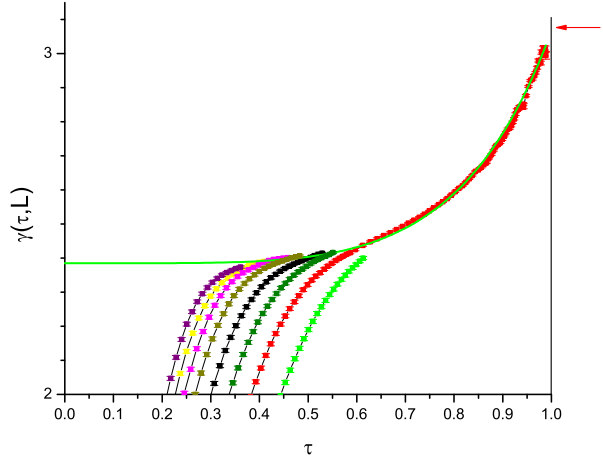}
  \caption{(Color online) The $4$d Laplacian temperature dependent
    effective exponent $\gamma(\tau)=\partial \chi(\beta,L)/ \partial
    \tau$ for $\beta_{c}= 0.662$. Symbol code as in Fig.~\ref{fig:1}. Green
    curve : ThL regime envelope fit.}\protect\label{fig:11}
\end{figure}

\begin{figure}
  \includegraphics[width=3.5in]{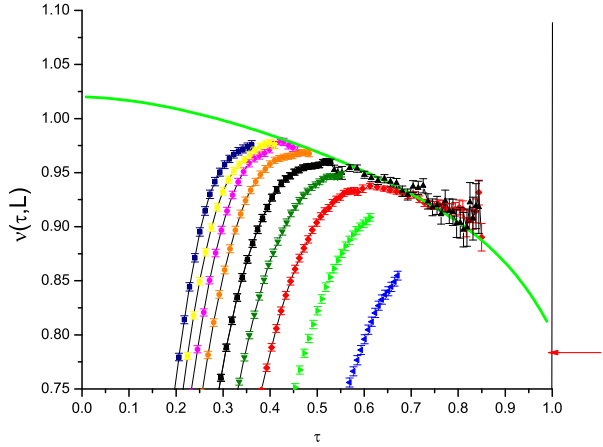}
  \caption{(Color online) The $4$d Laplacian effective exponent
    $\nu(\tau)=\partial (\xi(\beta,L)/\beta)/ \partial \tau$ with
    $\beta_{c}= 0.662$. Symbol code as in Fig.~\ref{fig:1}. Green curve : ThL
    regime data fit. }\protect\label{fig:12}
\end{figure}

\begin{figure}
  \includegraphics[width=3.5in]{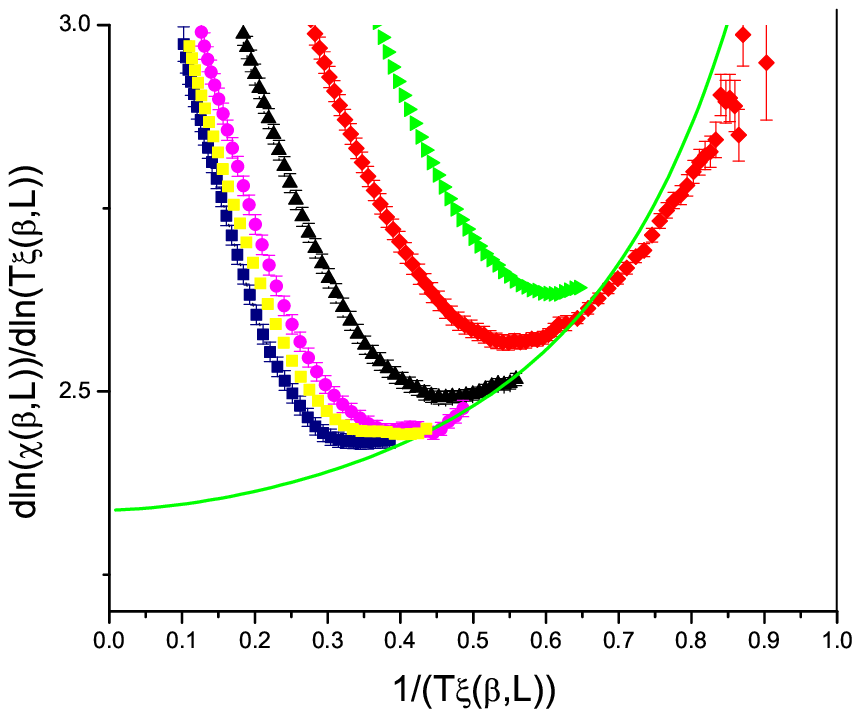}
  \caption{(Color online) The $4$d Laplacian effective exponent
    $2-\eta(\beta)=\partial\ln\chi(\beta,L)/\partial\ln(\xi(\beta,L)/\beta)$
    against $\beta/\xi(\beta,L)$.  Symbol code as in
    Fig.~\ref{fig:1}. Green envelope curve : ThL regime data fit and
    extrapolation. }\protect\label{fig:13}
\end{figure}

Simulation data were obtained for the Laplacian interaction ISG,
$P(J)\propto \exp(-|J|)$, in dimension $4$ with the same method and
analyses as for the Gaussian, and with $2^{13}$ samples for each size
$L$ up to $L=12$.  The FSS of the $W_{q}(\beta,L)$ data,
Fig.~\ref{fig:9}, showed negligible corrections to scaling, and the
other dimensionless parameters showed only very weak corrections. In
consequence $\beta_c$, $\nu$, and the critical values of the
dimensionless parameters could be estimated rather precisely,
Table~\ref{Table:II}.  The thermodynamic peak location data,
Fig.~\ref{fig:10}, extrapolate to a $\beta_{c}$ estimate which is
consistent with the FSS estimate. The susceptibility and correlation
length ThL measurements give estimates of the critical exponents
$\gamma$ and $\nu$ which are also in full agreement with the FSS and
pseudo-critical peak data, see Fig.~\ref{fig:11} and
Fig.~\ref{fig:12}. They also show weak correction term factors.

\begin{table}[htbp]
  \caption{\label{Table:II} Values of the $4$d Laplacian ISG FSS
    analysis fit parameters with standard errors. Dimensionless
    parameters $g(\beta,L)$, $W_{q}(\beta,L)$, $h(\beta,L)$ and
    $[\xi/L](\beta,L)$.}
  \begin{ruledtabular}
    \begin{tabular}{ccc}
      $g(\beta_c,\infty)$&$0.4747$&$0.0006$ \\
      $A(g)$&$-0.006$&$0.02$\\
      $B(g)$&$0.51$&$0.02$\\
      $\beta_{c}(g)$&$0.6226$&$0.0001$\\
      $\nu(g)$&$1.02$&$0.01$\\
      $W_{q}(\beta_c,\infty)$&$0.2385$&$0.0004$ \\
      $A(W_{q})$&$-0.15$&$0.01$\\
      $B(W_{q})$&$0.39$&$0.01$\\
      $\beta_{c}(W_{q})$&$0.6215$&$0.0010$\\
      $\nu(W_{q})$&$1.02$&$0.01$\\
      $h(\beta_c,\infty)$&$0.385$&$0.001$ \\
      $A(h)$&$0.060$&$0.015$\\
      $B(h)$&$0.48$&$0.01$\\
      $\beta_{c}(h)$&$0.6223$&$0.0004$\\
      $\nu(h)$&$1.026$&$0.010$\\
      $[\xi/L](\beta_c,\infty)$&$0.446$&$0.001$ \\
      $A(\xi/L)$&$0.65$&$0.10$\\
      $B(\xi/L)$&$0.38$&$0.01$\\
      $\beta_{c}(\xi/L)$&$0.6220$&$0.0003$\\
      $\nu(\xi/L)$&$1.03$&$0.01$\\
    \end{tabular}
  \end{ruledtabular}
\end{table}

\begin{figure}
  \includegraphics[width=3.5in]{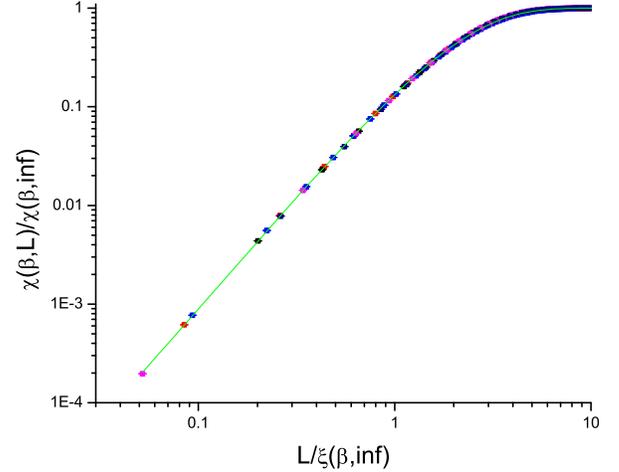}
  \caption{(Color online) The $4$d Laplacian Privman-Fisher $\chi$ FSS
    plot, $\chi(\tau,L)/\chi(\tau,\infty)$ against
    $L/\xi(\tau,\infty)$ where $\chi(\tau,\infty)$ and
    $\xi(\tau,\infty)$ are fit and extrapolation values from the
    $\gamma(\tau)$ and $\nu(\tau)$ analyses. Symbol code as in
    Fig.~\ref{fig:1}. Curve : fit (see text)}\protect\label{fig:14}
\end{figure}

\begin{figure}
  \includegraphics[width=3.5in]{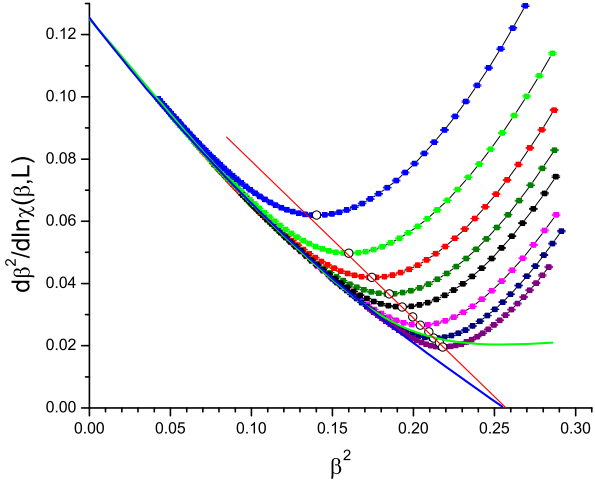}
  \caption{(Color online) The $4$d bimodal derivatives $\partial
    \beta^2/ \partial \ln\chi(\beta,L)$ as a function of
    $\beta^2$. Symbol code as in Fig.~\ref{fig:1}. Green curve : ThL
    derivative from the 15 exact term HTSE series \cite{daboul:04}.
    Blue curve : ThL regime data fit and extrapolation.  Open circles
    : minima locations for each $L$, straight line : extrapolation to
    $\beta_{c}^2$. }\protect\label{fig:15}
\end{figure}

The overall conclusions from the analyses for the Laplacian are :
$\beta_{c} = 0.6221(5)$, with critical exponents $\nu = 1.026(5)$,
$\gamma = 2.385(10)$, $\eta = -0.32(4)$ and critical values for the
dimensionless parameters $g(\beta_{c},\infty) = 0.4747(6)$,
$W_{q}(\beta_c,\infty) = 0.2385(4)$, $h(\beta_{c},\infty) =0.385(1)$,
and $[\xi/L](\beta_c,\infty)=0.446(1)$.  For all the parameters the
correction exponent $\theta$ or $\omega$ values are large, we have
used $\omega=3\pm 1$ for the fits and error estimates. As these are
effective exponents the values are not the same for all the
parameters. The overall fit to the $\chi(\tau,L)$ data with the
leading correction term is $\chi(\tau,\infty) =
1.14\tau^{-2.385}(1-0.12\tau^5)$.  The overall fit to the
$\xi(\tau,L)$ data with a leading and a high order correction term is
\begin{equation}
  \xi(\tau,\infty) =
  0.90\,\beta\,\tau^{-1.02} (1+0.1\,\tau^{1.7}+0.007\,\tau^{10})
\end{equation}  
The Privman-Fisher FSS fit Eq.~\eqref{PFansatz}, Fig.~\ref{fig:14}, is
\begin{equation}
  \frac{\chi(\tau,L)}{\chi(\tau,\infty)}= \left(1 -
  \exp\left[-0.38\left(\frac{L}{\xi(\tau,\infty)}\right)^{\frac{2.3}{1.75}}\right]\right)^{1.75}
\end{equation}

\section{The bimodal ISG in dimension $4$}

For the $4$d bimodal ISG, from early simulation measurements up to $L
= 10$ a critical temperature $\beta_{c}=0.493(7)$ (i.e. $\beta_{c}^2 =
0.243(7)$) was estimated \cite{marinari:99} using the Binder parameter
crossing point criterion. However, finite size corrections to scaling
were not allowed for. The exponent estimates were $\nu = 1.0(1)$ and
$\eta = -0.30(5)$. Extensive domain wall free energy measurements to
$L = 10$ gave an estimate $\beta_{c}=0.50(1)$ (i.e. $\beta_{c}^2 =
0.25(1)$) \cite{hukushima:99}.  Inspection of the raw data
\cite{hukushima} shows strong finite size corrections; extrapolation
to larger $L$ leads to an infinite size limit $\beta_{c}$ definitely
greater than $0.50$. The HTSE critical temperature and exponent
estimates are \cite{daboul:04} $\beta_{c}^2 = 0.26(2)$
(i.e. $\beta_{c} = 0.51(2)$), $\gamma = 2.5(3)$ and a leading
confluent correction exponent $\theta \sim 1.5$.

From bimodal $4$d simulations with impressive numbers of samples up to
$L=16$ and to a maximum inverse temperature $\beta = 0.5025$, Ba\~nos
{\it et al.} \cite{banos:12} give estimates $\omega = 1.04(10)$, $\nu
= 1.068(7)$ (so $\theta = \omega\nu = 1.11(12)$), $\eta = -0.320(13)$
(so $\gamma = (2-\eta)\nu = 2.48(3)$) and $\beta_c =0.5023(6)$, all
from a FSS analysis with $t$ as scaling variable. We can note that the
inverse temperatures for the $\left(U_{4}(\beta,L),
U_{4}(\beta,2L)\right)$ and $\left([\xi/L](\beta,L),
[\xi/L](\beta,2L)\right)$ crossing points should scale linearly with
$1/L^{\omega+1/\nu}$, Eq. 30 of \cite{banos:12}. However, it can be
seen that in the crossing point plot, Fig.~9 of Ref.~\cite{banos:12},
with the authors' preferred values of $\omega$ and $\nu$ the scaling
against $1/L^{\omega+1/\nu}$ is far from being linear. Ba\~nos {\it et
  al.} obtain fits by discarding the lower $L$ points or by
introducing strong higher order terms. An alternative explanation
could be that despite the precautions taken complete equilibration has
not quite been achieved for the largest $\beta$ measurements at the
largest size $L=16$ (where equilibration is the most difficult), so
the $L=16$ to $L=8$ crossing points could be dropped. In this case the
remaining data points appear more consistent and indicate a rather
larger critical inverse temperature $\beta_{c}$ and a rather larger
$\omega+1/\nu$. Our independent data reported below are compatible
with these modified values.

\begin{figure}
  \includegraphics[width=3.5in]{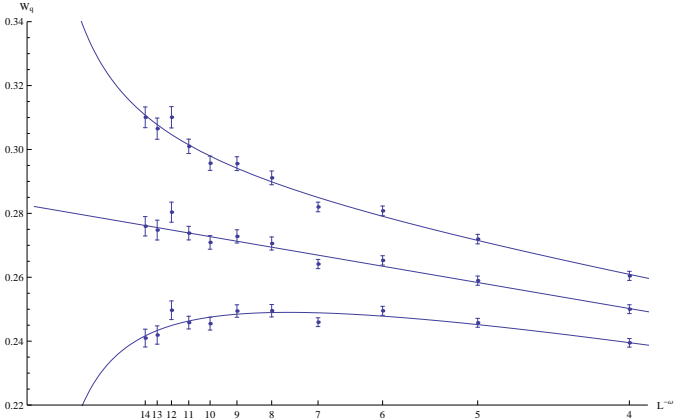}
  \caption{(Color online) The $4$d bimodal $W_{q}$ finite size
    scaling. Data and fit curves : upper $\beta=0.51032$, center
    $\beta=0.50532$, lower $\beta=0.50032$.  Fit value $\omega =
    1.3$. See text and Table~\ref{Table:III}.}\protect\label{fig:16}
\end{figure}

We have made measurements in the critical region of the standard
finite size Binder cumulant $g(\beta,L)$, the dimensionless parameters
$W_{q}(\beta,L)$ and $h(\beta,L)$, and the normalized correlation
length $[\xi/L](\beta,L)$. The present data are for all $L$ from $3$
to $14$ and span the estimated inverse critical temperature
$\beta_c$. From fits to the $g(\beta,L)$, $W_{q}(\beta,L)$, and
$h(\beta,L)$ we obtain FSS estimates for the critical temperature
$\beta_c$ and the exponent $\nu$ together with the dimensionless
critical values $g(\beta_{c},\infty)$, $W_{q}(\beta_{c},\infty)$, see
Fig.~\ref{fig:16}, and $h(\beta_{c},\infty)$.  The correlation length
ratio $[\xi/L](\beta,L)$ has much larger corrections to scaling than
the other observables, so the data for this parameter was not readily
exploitable.  As explained above rather than following the traditional
crossing point analysis, the raw dimensionless parameter data are
fitted directly to Eq.~\eqref{FSS} at fixed temperatures in the
critical region.  Consistent independent fits could be made to the
$g(\beta,L)$, $W_{q}(\beta,L)$ and $h(\beta,L)$ results for the data
from $L=4$ to $L=14$. (To minimize higher order corrections we have
eliminated $L=3$ from the fitting process). Fits were made fixing the
correction exponent at three alternative values, $\omega = 1.0$, $1.3$
and $1.6$. The fit values with error bars which include the effect of
the assumed uncertainty in $\omega$ are given in
Table~\ref{Table:III}.

\begin{table}[htbp]
  \caption{\label{Table:III}  Values of the $4$d bimodal ISG FSS
    analysis fit parameters with standard errors. Dimensionless
    parameters $g(\beta,L)$, $W_{q}(\beta,L)$, $h(\beta,L)$ and
    $[\xi/L](\beta,L)$.}
  \begin{ruledtabular}
    \begin{tabular}{ccc}
      $g(\beta_c,\infty)$&$0.5258$&$0.0055$ \\
      $A(g)$&$-0.2089$&$0.026$\\
      $B(g)$&$0.7485$&$0.037$\\
      $\beta_{c}(g)$&$0.5058$&$0.0007$\\
      $\nu(g)$&$1.08$&$0.03$\\
      $W_{q}(\beta_c,\infty)$&$0.2789$&$0.0031$ \\
      $A(W_{q})$&$-0.1820$&$0.015$\\
      $B(W_{q})$&$0.5883$&$0.021$\\
      $\beta_{c}(W_{q})$&$0.5045$&$0.0005$\\
      $\nu(W_{q})$&$1.07$&$0.02$\\
      $h(\beta_c,\infty)$&$0.435$&$0.010$ \\
      $A(h)$&$-0.215$&$0.025$\\
      $B(h)$&$0.7199$&$0.023$\\
      $\beta_{c}(h)$&$0.5056$&$0.0006$\\
      $\nu(h)$&$1.09$&$0.02$\\
      $[\xi/L](\beta_c,\infty)$&$0.4780$&$0.0028$ \\
      $A(\xi/L)$&$-0.2925$&$0.014$\\
      $B(\xi/L)$&$0.5111$&$0.019$\\
      $\beta_{c}(\xi/L)$&$0.5036$&$0.0005$\\
      $\nu(\xi/L)$&$1.02$&$0.02$\\
    \end{tabular}
  \end{ruledtabular}
\end{table}

Estimates for the important critical values of the dimensionless
parameters for $4$d bimodal ISG are not quoted explicitly in
Ref.~\cite{banos:12}, but a limit $U_{4}(\beta_{c}) < 2.00$
i.e. $g(\beta_{c},\infty)> 0.50$ can be read off the appropriate
figure.  From the present bimodal analysis,
$g(\beta_{c},\infty)=0.526(6)$ (or $U_{4}(\beta_{c},\infty)
=1.948(12)$), $W_{q}(\beta_{c},\infty)= 0.279(3)$ and
$h(\beta_{c},\infty)= 0.435(10)$.

\begin{figure}
  \includegraphics[width=3.5in]{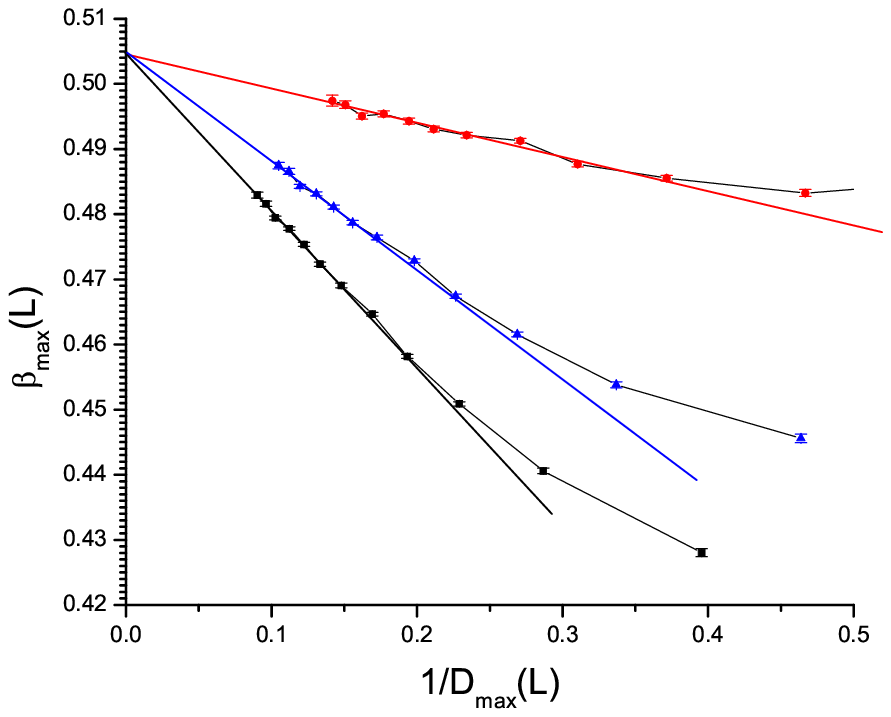}
  \caption{(Color online) The $4$d bimodal thermodynamic derivative
    peaks; peak location $\beta_{\max}(L)$ as function of inverse peak
    heights $1/D_{\max}(L)$ where $D_{\max}(L) =\max \partial U(\beta,L)/
      \partial \beta$.  Observables $U$: red circles $W_{q}$,
    blue triangles $h$, black squares $g$.
    Straight line fits indicate extrapolations to $\beta_c$ (see
    text).}\protect\label{fig:17}
\end{figure}

Thermodynamic derivative peak data are shown in Fig.~\ref{fig:17}.
The straight line extrapolations of peak locations $\beta_{\max}(L)$
against the inverse peak strengths $1/\max[\partial
  U(\beta,L)/\partial\beta]$ for observables $U(\beta,L)$ show
consistently $\beta_{\max}(L)$ values tending to a $\beta_{c} =
0.5050(5)$ in the infinite $L$ limit.  Log-log plots of $\max[\partial
  U(\beta,L)/\partial\beta]$ versus $L$ tend to straight lines with
limiting slopes $1/\nu$ corresponding to $\nu = 1.12(2)$. It should be
again underlined that these estimates for $\nu$ are independent of the
$\beta_{c}$ estimates.

\begin{figure}
  \includegraphics[width=3.5in]{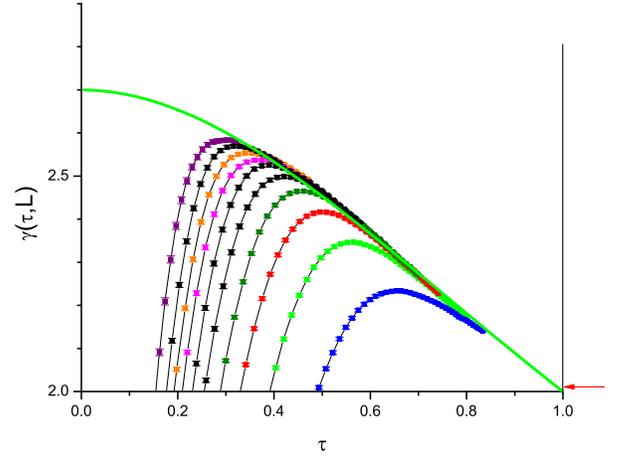}
  \caption{(Color online) The $4$d bimodal temperature dependent
    effective exponent $\gamma(\tau)=\partial \chi(\beta,L)/ \partial
    \tau$ for $\beta_{c}= 0.505$. Symbol code as in Fig.~\ref{fig:1}. Green
    curve : HTSE data points. Blue curve : ThL regime data
    fit. }\protect\label{fig:18}
\end{figure}
\begin{figure}
  \includegraphics[width=3.5in]{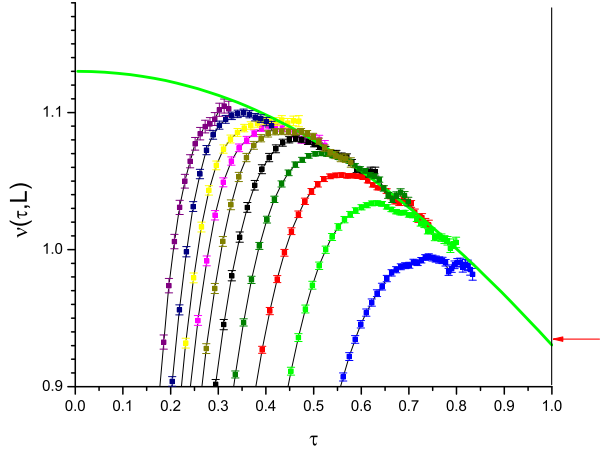}
  \caption{(Color online) The $4$d bimodal effective
    $\nu(\tau)=\partial (\xi(\beta,L)/\beta)/ \partial \tau$ for
    $\beta_{c}= 0.505$. Symbol code as in Fig.~\ref{fig:1}. Green curve : ThL
    regime data fit. }\protect\label{fig:19}
\end{figure}
\begin{figure}
  \includegraphics[width=3.5in]{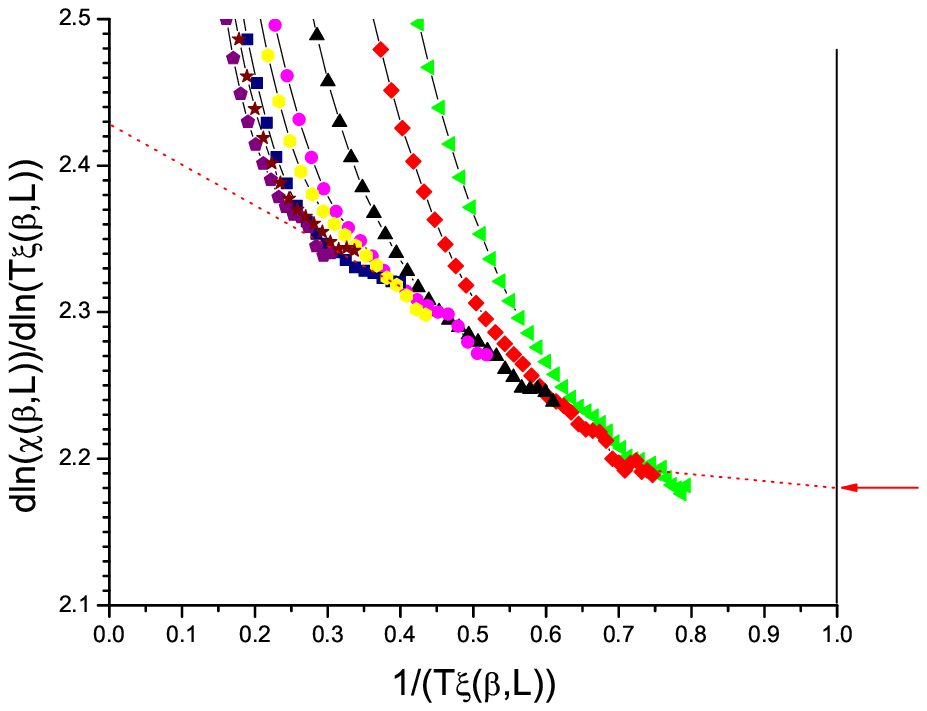}
  \caption{(Color online) The $4$d bimodal effective exponent
    $2-\eta(\beta)=\partial\ln\chi(\beta,L)/\partial\ln(\xi(\beta,L)/\beta)$
    against $\beta/\xi(\beta,L)$.  Symbol code as in
    Fig.~\ref{fig:1}. Dotted envelope curve : ThL regime data fit and
    extrapolation. }\protect\label{fig:20}
\end{figure}

We show in Fig.~\ref{fig:18} and Fig.~\ref{fig:19} $\gamma(\tau,L)$
and $\nu(\tau,L)$ plots for $4$d binomial ISG using as the inverse
critical temperature $\beta_{c} =0.505$ as estimated from the FSS and
thermodynamic peak analyses. Satisfactory fits can be obtained with
$\nu=1.14(2)$, $\theta \sim 1.8$, and $\gamma= 2.76(3)$ with
prefactors $a_{\xi} =0.10, a_{\chi} = 0.69$ and a weak high order
correction term for $\gamma(\tau)$, $y \sim 8, b_{\chi} \sim
-0.01$. There is consistency between the FSS exponent estimates and
the ThL estimates.


In Ref~\cite{banos:12}, the FSS estimate of the bimodal ISG critical
temperature is $\beta_c =0.5023(6)$.  Fixing $\beta_c$ at this value,
a ThL $\gamma(\tau)$ fit leads to an estimate $\gamma =2.60(3)$, still
considerably above the Gaussian value. However, the fit requires a
correction exponent $\theta \sim 2.0$, corresponding to $\omega \sim
1.8$, which is considerably higher than the Ref.~\cite{banos:12} FSS
value $\omega = 1.04(10)$.

\begin{figure}
  \includegraphics[width=3.5in]{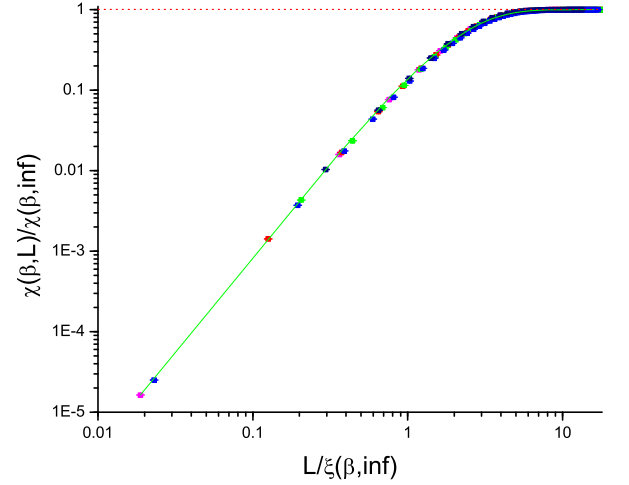}
  \caption{(Color online) The $4$d bimodal Privman-Fisher $\chi$ FSS
    plot, $\chi(\tau,L)/\chi(\tau,\infty)$ against
    $L/\xi(\tau,\infty)$ where $\chi(\tau,\infty)$ and
    $\xi(\tau,\infty)$ are fit and extrapolation values from the
    $\gamma(\tau)$ and $\nu(\tau)$ analyses. Symbol code as in
    Fig.~\ref{fig:1}. Curve : fit (see text)}\protect\label{fig:21}
\end{figure}

Finally Privman-Fisher ansatz plots can be made up for all the
data. Assuming $\beta_{c}=0.505$ and the fit parameters from the
$\gamma(\tau,L)$ and $\nu(\tau,L)$ plots, the Privman-Fisher scaling
is shown in Fig.~\ref{fig:21}.  It can be seen that the $\chi$ scaling is
excellent on the scale of the figure over the entire range of
paramagnetic temperatures and of sizes. The scaling curve can be
fitted accurately by the explicit form used to fit the simple cubic
Ising ferromagnet data, Eq.~\eqref{PFansatz} with $\eta=-0.40 $,
$a=1.95$, and $b=0.43$ for the $\chi$ scaling plot.  For the $\xi$
scaling plot a good large $L$ fit is obtained using the ansatz
Eq.~\eqref{PFxiansatz} with $a = 0.74$, and $b= 0.35$ with indications
of finite size corrections which we have not attempted to analyse.  It
can be noted that both for $\chi$ and $\xi$ the values of the fit
parameters (which should be characteristic of a universality class)
obtained for the bimodal, Gaussian and Laplacian ISGs are not quite
identical.

The present critical exponents $\gamma$ for both the bimodal and
Gaussian ISGs are consistent with, but are more accurate than, the
HTSE estimates, principally because the uncertainty in $\beta_{c}^2$
is considerably reduced thanks to the FSS and thermodynamic analyses
of the simulation data.  The $4$d bimodal $\gamma$ and $\nu$ exponent
estimates can be compared to the values found above for the $4$d
Gaussian, and with the published estimates Gaussian and diluted
bimodal systems \cite{jorg:08}.  The critical exponents of the $4$d
bimodal ISG are well above those of the $4$d Gaussian and diluted
bimodal ISGs.

\section{Comparisons between models}

The bimodal, Gaussian, Laplacian and diluted bimodal infinite size
limit critical values for the Binder cumulant can be compared:
\begin{itemize}
\item bimodal $g(\beta_c,\infty)= 0.526(5)$ (or $> 0.500$ \cite{banos:12}),
\item Gaussian $g(\beta_c,\infty)= 0.484(3)$ (or $0.470(5)$ \cite{jorg:08}),
\item Laplacian $g(\beta_c,\infty)= 0.4747(6)$
\item diluted $g(\beta_{c},\infty)=0.472(2)$ \cite{jorg:08}.
\end{itemize}

The value estimated for the bimodal model is strikingly different from
those for the other models. Given the high precision of the other
estimates it would appear that they are not quite identical to each
other either.  For the other dimensionless parameters (measurements
have been made only in the present work)

\begin{itemize}
\item bimodal $W_{q}(\beta_c,\infty)= 0.279(3)$
\item Gaussian $W_{q}(\beta_c,\infty)= 0.248(3)$
\item Laplacian $W_{q}(\beta_c,\infty)= 0.2385(4)$
\end{itemize}
and
\begin{itemize}
\item bimodal $h(\beta_c,\infty)= 0.435(10)$
\item Gaussian $h(\beta_c,\infty)= 0.394(4)$,
\item Laplacian $h(\beta_c,\infty)= 0.385(1)$.
\end{itemize}

The distinction between the bimodal value and the other two values is
again very significant.  As the critical limit values of dimensionless
parameters are characteristic of a university class, even without
knowing the critical exponents these model-to-model differences
already clearly demonstrate that the $4$d binomial ISG is not in the
same universality class as the other three $4$d ISG models.

The critical exponents from a combination of FSS, thermodynamic peak,
and ThL information (when available) are
\begin{itemize}
\item bimodal $\gamma(\beta_c,\infty)= 2.70(4)$,
\item Gaussian $\gamma(\beta_c,\infty)= 2.44(4)$,
\item Laplacian $\gamma(\beta_c,\infty)= 2.40(2)$,
\item diluted $\gamma(\beta_c,\infty)= 2.33(6)$ \cite{jorg:08}
\end{itemize}
and
\begin{itemize}
\item bimodal $\nu(\beta_c,\infty)= 1.13(1)$,
\item Gaussian $\nu(\beta_c,\infty)= 1.030(5)$,
\item Laplacian $\nu(\beta_c,\infty)= 1.020(5)$
\item diluted $\nu(\beta_c,\infty)= 1.025(15)$ \cite{jorg:08}
\end{itemize}

The two critical exponents $\gamma$ and $\nu$ for the bimodal
interaction model are both much higher than the values for the
Gaussian, Laplacian, and diluted bimodal interaction models.

\section{Conclusions}

Simulations on the $4$d Gaussian, Laplacian and bimodal ISGs up to
size $L=12$ and $L=14$ respectively are first analysed in the critical
temperature range to obtain estimates for the critical inverse
temperatures $\beta_{c}$, together with FSS estimates for the critical
values for dimensionless parameters $g(\beta_{c},\infty)$,
$W_{q}(\beta_{c},\infty)$ of Eq.~\eqref{Wqdef} and
$h(\beta_{c},\infty)$ of Eq.~\eqref{hdef}, and for the anomalous and
correlation length critical exponents $\eta$ and $\nu$.  The thermal
derivative peak method has been used to obtain complementary estimates
for $\beta_c$, and independent estimates of the critical exponent
$\nu$ which are not dependent on the $\beta_c$ estimates. The critical
temperatures $\beta_{c}$ are consistent with the FSS estimates and are
in full agreement with, but are more precise than, the estimates from
high temperature scaling expansions alone \cite{daboul:04}.  They also
improve on previous FSS numerical estimates \cite{jorg:08,banos:12}.

We spell out in detail the procedure used for scaling over the whole
paramagnetic temperature range with the appropriate ISG scaling
variable, $\tau = 1-(\beta/\beta_{c})^2$, together with scaling
expressions which include the correction terms, Eqs.~\eqref{chiISG}
and \eqref{xiISG}.

With the estimates for $\beta_{c}$ in hand, data for the temperature
dependent effective exponents $\gamma(\tau)$ and $\nu(\tau)$ in the
thermodynamic limit (ThL) regime were fitted and extrapolated to
obtain critical exponent $\gamma$, $\nu$ and $\theta$ estimates
together with the strengths of the correction terms. There is
consistency between the FSS, thermodynamic derivative peak, and ThL
estimates for each model. Overall Privman-Fisher scalings with the
estimated critical parameters, covering the whole paramagnetic
temperature range and all sizes $L$, validate the analysis.

The critical values of the dimensionless parameters and the critical
exponents are characteristic of a universality class. For each of the
recorded observables the values for the $4$d bimodal ISG are quite
different from those of the $4$d Gaussian ISG and the $4$d Laplacian
(or from those of the $4$d diluted bimodal ISGs \cite{jorg:08}),
showing that these systems lie in different universality classes.  It
can be concluded that for ISG transitions in dimension $4$ at least,
and probably more generally \cite{lundow:13a,lundow:13b}, the critical
parameters depend on the form of the interaction distribution, so the
standard RGT universality rules do not apply.

\section{Acknowledgements}
We are very grateful to Koji Hukushima for important comments and
communication of unpublished data. We thank Amnon Aharony for
constructive criticism. The computations were performed on resources
provided by the Swedish National Infrastructure for Computing (SNIC)
at the High Performance Computing Center North (HPC2N) and Chalmers
Centre for Computational Science and Engineering (C3SE).


\begin{thebibliography}{20}
\bibitem{notebeta} We will use the inverse temperature $\beta = 1/T$ or its square throughout.
\bibitem{gauss13} The Gaussian series was calculated to $13$ terms but explicit evaluation shows that
there is a numerical error in the final term.
\end{thebibliography}
\end{document}